\documentclass[ACS,STIX2COL]{WileyNJD-v2}

\usepackage{makecell}

\articletype{Research Article}%

\begin{document}

\title{Causal Discovery to Understand Hot Corrosion}

\author[1]{Akhil Varghese}

\author[1]{Miguel Arana-Catania*}

\author[1]{Stefano Mori}

\author[1]{Adriana Encinas-Oropesa}

\author[1]{Joy Sumner}

%\authormark{Akhil Varghese \textsc{et al}}

\address[1]{\orgname{Cranfield University}, \country{UK}}

\corres{*Miguel Arana-Catania. \email{miguel.aranacatania@cranfield.ac.uk}}

%\presentaddress{This is sample for present address text this is sample for present address text}

\abstract{Gas turbine superalloys experience hot corrosion, driven by factors including corrosive deposit flux, temperature, gas composition, and component material. The full mechanism still needs clarification and research often focuses on laboratory work. As such, there is interest in causal discovery to confirm the significance of factors and identify potential missing causal relationships or co-dependencies between these factors. The causal discovery algorithm Fast Causal Inference (FCI) has been trialled on a small set of laboratory data, with the outputs evaluated for their significance to corrosion propagation, and compared to existing mechanistic understanding. FCI identified the salt deposition flux as the most influential corrosion variable for this limited dataset. However, HCl was the second most influential for pitting regions, compared to temperature for more uniformly corroding regions. Thus FCI generated causal links aligned with literature from a randomised corrosion dataset, while also identifying the presence of two different degradation modes in operation.}

\keywords{Gas Turbine superalloys, Hot Corrosion, Causal Inference, Causal Discovery Method, FCI Algorithm, Kernel-based
Conditional Independence Test}



\maketitle

\section{Introduction}
\label{bkm:Ref144506525}
Gas turbines are used to generate power and provide thrust \citep{Harman1981}. In 2023 between 26\% and 37\% of UK electricity was generated from natural gas combustion, with gas turbines now shifting towards zero emission fuels such as H$_2$ or NH$_3$, to meet UN goals. In order to increase the efficiency, gas turbines operate in a combined cycle with steam turbines, and are required to operate at higher temperatures and pressures,  generating a more challenging environment for the materials used for their manufacture. 
Furthermore, contaminants such as sulphates, halides, and chlorides, contained in the fuel, together with salt
impurities from the air, can create a highly corrosive environment \citep{eliaz2002hot}.
Deposition of such contaminants on to the gas turbine's component blades and vanes gives
rise to a corrosion mechanism called ``hot corrosion", which adversely affects the service life of the gas turbines. Depending on the operating temperature conditions inside the gas turbine, Type-I or Type-II hot
corrosion can occur \citep{pettit2011hot}. Type-I and Type-II hot corrosion are temperature-dependent and deposit-induced
corrosion mechanisms that occur across the approximate temperature ranges of 850-950{\textdegree}C and 650-800{\textdegree}C, respectively
\citep{eliaz2002hot}. Melting temperatures of the salt contaminants influence both these types of accelerated corrosion
mechanisms but factors such as a high partial pressure of SO$_3$ are crucial for Type-II hot corrosion to occur \citep{eliaz2002hot}. 
While it is known that factors like operating temperatures, melting temperatures of salt contaminants, the deposition rate of the flux on gas turbine surface, partial pressures of gas contaminants, and material and gas compositions lead to hot corrosion in gas turbines, the exact variation of the underlying mechanism with changing parameters needs clarification.

Moreover, the degree of co-dependence and independence between these factors will help in understanding the influence that each factor has on
hot corrosion. Hence, Causal Discovery \citep{pearl2009causality,glymour2019review} has been introduced to understand the
causal relationships between the different corrosion factors from a statistical approach. These data-driven, causal
relationships will help in understanding the underlying causal relationships of each corrosion variable, and thus clarifying the physical corrosion mechanism. Such findings from
this early-stage study can also help in designing future hot corrosion experiments, both in terms of the variables and their values, and can become the foundation upon which
a predictive maintenance model can be developed.

\subsection{Hot Corrosion}
\label{bkm:Ref144506786}\label{bkm:Ref144507318}Hot corrosion is a type of material degradation which is induced by the deposition of contaminants contained in the exhaust stream of a gas turbine \citep{muktinutalapati2011materials}. These deposits are usually formed by alkali compounds that can increase the corrosion rates if in molten state \citep{pettit2011hot}. Some of the contaminants come from the air intake (such as sodium, potassium, calcium), while others come from the fuel. 
The amount of deposition influences the corrosion rate of the materials in such environments \citep{sumner2013type}. Based on the operating temperatures and the type of contaminant attack, hot corrosion can be further divided into Type-I
and Type-II hot corrosion:

\subsubsection{Type-I Hot Corrosion (High-temperature Hot
Corrosion)}
\label{bkm:Ref144509428}Type-I hot corrosion is also called high-temperature hot corrosion (HTHC) because it occurs within the temperature range of 850-950{\textdegree}C \citep{eliaz2002hot}. Various compounds can form, for example, Sodium from the ingested air and sulphur from the fuel could combine to form sodium sulphate (Na$_2$SO$_4$), during the combustion cycle of the gas turbine
\citep{schilke2004advanced} \citep{Mori2023}. The sodium sulphate could then condense onto the colder blading, and form a liquid (molten) salt, which initiates  Type-I hot corrosion. In the presence of further impurities, such as  NaCl from the industrial or marine atmosphere combines or K from sea droplets  \citep{Mori2023}, the formation of mixtures with lower melting points could occur which extend the temperature range of Type-I hot corrosion \citep{reed2008superalloys}. 

\begin{figure}[!ht]
\noindent
\centerline{ \includegraphics[width=0.5\textwidth]{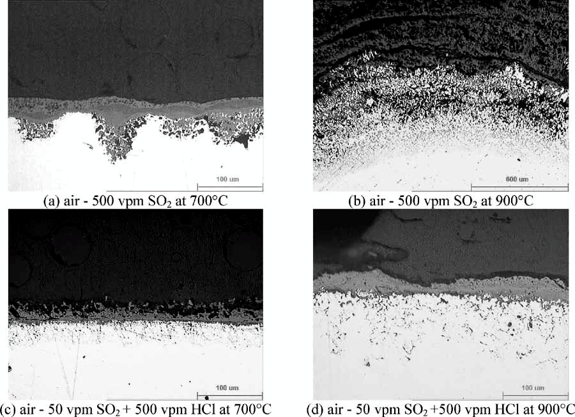} }
\caption{\label{bkm:Ref144506990}Optical micrographs of SC$^2$-B exposed for 500h with a flux of 5 µg/cm$^2$/h. The picture shows pitting attack and scale formation at 700 and 900 \textdegree C \citep{encinas2005study}.   }
\end{figure}

Using a fluxing reaction, these eutectic mixtures dissolve the protective layer of the superalloy and attack the
base material. These mixtures shorten the incubation
period. Type-I hot corrosion is characterised by internal sulfidation, depletion of
protective layers formed by chromium or aluminium, and leads to severe metal loss from the base material, impacting the gas turbine life. Sulfidation can be
seen in Figure \ref{bkm:Ref144506990}(b) and (d).

\subsubsection{Type-II Hot Corrosion (Low-temperature Hot
Corrosion)}
\label{bkm:Ref144509436}\label{bkm:Ref144509601}Type-II hot corrosion, also called low-temperature hot corrosion
(LTHC), occurs at temperatures between 650-800{\textdegree}C \citep{pettit2011hot}. As in the case of Type-I hot corrosion, the formation of a molten deposit is also involved in the mechanism controlling the corrosion behaviour. However, the mechanism occurs at temperatures below the melting temperature of many pure salts. In Type-II hot corrosion the Na$_2$SO$_4$ forms mixtures with metallic inorganic compounds \citep{eliaz2002hot}, which lowers the combined melting temperature and initiates the corrosion process. These inorganic metal compounds are formed by SO$_3$ present in the combustion gas and the metal \citep{reed2008superalloys}, making Type-II hot corrosion not only a function of the temperature, but also a function of the partial pressure of the SO$_3$ gas \citep{pettit2011hot}. Type-II hot corrosion is characterised by pitting with
localized failure as seen in Figure \ref{bkm:Ref144506990}(a) \citep{eliaz2002hot}. The cause of pitting initiation is currently under consideration and has been linked to a wide range of factors including grain boundaries, precipitates, gas environment and the salt deposit.

\subsection{Gas Turbine Materials}
\label{bkm:Ref144506808} To increase efficiency, higher turbine operating temperatures and pressures are required. This has driven the formulation and selection of materials to be used, which need to have \citep{reed2008superalloys}:
\begin{itemize}
\item high mechanical strength at temperatures close to the melting point
\item high creep resistance 
\item high corrosion resistance
\end{itemize}

For these reasons ``superalloys" with better mechanical properties than conventional alloys have been developed. These alloys are  nickel-, iron-nickel and
cobalt-based  \citep{donachie2002superalloys}, with other elements in solution, such as chromium and aluminium, which preferentially oxidise to form a thin protective oxide layer to provide resistance to corrosion and oxidation \citep{donachie2002superalloys}. 

To ensure resistance to hot corrosion, a minimum amount of chromium is needed, for instance,
22\% chromium is present in Co-base Haynes 188 superalloy, which provides high fatigue strength and strong resistance
to hot corrosion \citep{kool1996current}. In comparison, even though nickel-base Haynes 214 contains 16\% chromium, it provides
high oxidation resistance to temperatures above 900\textdegree C due to the presence of 4.5\% aluminium \citep{donachie2002superalloys}.

Microstructure can also be controlled with commercially used single crystal superalloys like CMSX-11C and SC-16 providing increased resistance to hot corrosion due
to the presence of more than 12\% chromium \citep{muktinutalapati2011materials}.

\subsubsection{Protective Coatings}
\label{bkm:Ref144506823}\label{bkm:Ref144507326}
\begin{figure}[!ht]
\noindent
 \centerline{\includegraphics[width=0.5\textwidth]{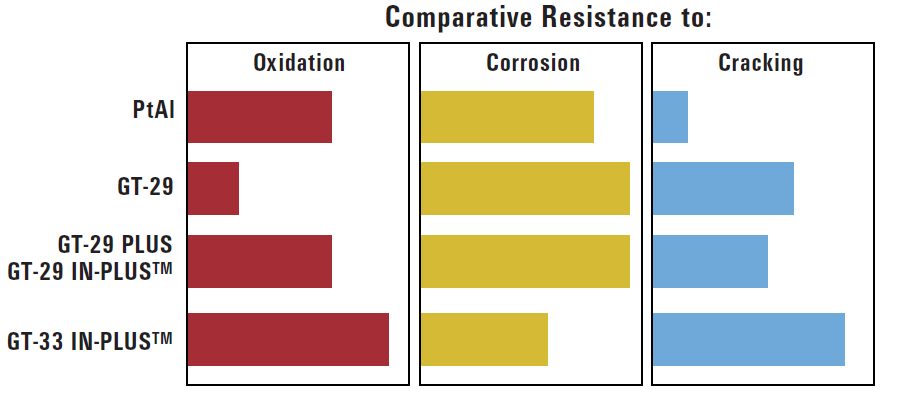} }
\caption{\label{bkm:Ref144507264}Comparison of resistance performance between
platinum-aluminide, and overlay coatings \citep{schilke2004advanced} }
\end{figure}

Despite the more recent advancements, most of the superalloys are not able to provide the desired lifetime in all conditions within gas turbines. Thus, for specific areas of the turbines, coatings are required \citep{muktinutalapati2011materials,donachie2002superalloys}. Aluminide diffusion
coatings, overlay coatings and thermal barrier coatings are the most used types of coatings for gas turbine applications
\citep{goward1998progress}. Platinum-modified aluminide (Pt-Al) coatings like RT-22 and CN-91 are widely used \citep{vialas2006substrate} due to their great resistance to Type-I and Type-II hot corrosion \citep{wu1984role}. On the other hand, overlay coatings provide excellent oxidation and hot corrosion resistance due to their ability to form alumina and chromia scales \citep{pomeroy2005coatings}. Figure \ref{bkm:Ref144507264} shows the degradation resistance of a platinum-aluminide diffusion coating versus three
overlay coatings \citep{schilke2004advanced}. 

\subsubsection{Hot Corrosion summary}
Hot corrosion is dependent on different factors including the chemistry of the corroding material, temperature, partial pressure of contaminants and chemistry of the contaminants. For this reason, it is not straightforward to understand the connection between the different parameters and the rates of corrosion. Understanding the cause-effect relationship between the corrosion factors discussed can help in analysing materials' degradation in gas turbines. The discovery of cause-effect relationships can be done through the use of causal discovery techniques, which  can
investigate the links between these corrosion factors. 

\subsection{Causal Discovery}
\label{bkm:Ref144506832}\label{bkm:Ref144507340}

Causal analysis techniques have been prominently used in the fields of
engineering, medicine, and economics \citep{pearl2009causality}. Manufacturing, however, has yet to fully embrace causal discovery
methods when compared to the previous fields, and thus there is limited work on hot corrosion using these techniques \citep{vukovic2022causal}. 

One example of the use of causal discovery methods was applied to detect strong relationships in degradation data \citep{feng2022degradation}. A neural
network was used to assess the degradation state of some equipment. The causal discovery method FCI was used to create a model of the relationship between the variables. The variables responsible
for the degradation state found using FCI, were then fed into Long Short Term Memory (LSTM) neural networks for the
subsequent assessment. In this manner, the data-driven model was trained, and its interpretability
was improved.

In a second example \citep{chen2022introducing}, the method was applied to the design process of energy-efficient buildings. The authors  applied the causal discovery algorithm Greedy
Equivalence Search (GES) to the variables that potentially affect building design. Creating such a causal
framework is expected to allow designers, developers, and construction workers to inspect and continuously improve their own
designs and construction methodology.

In the final example reported here, causal discovery algorithms were used on an Alzheimer's disease dataset \citep{shen2020challenges}. The study compared between two causal discovery algorithms and an existing standard graph on Alzheimer's Disease, which was formed using literature and prior experience. FCI and Fast Greedy Equivalence Search were the methods used to form causal graphs, with initial causal graphs formed purely based on observational datasets i.e., without prior subject knowledge. Subsequently, `background knowledge' was added to the algorithms and the changes in average accuracy, recall and precision were compared with the former results. These causal graphs were later validated
based on the existing standard graph and the discovered graphs were found to be very close.

This research aims to understand the degree of influence, independence and co-dependence of several hot corrosion variables
causing material degradation in a gas turbine setting using causal discovery techniques. 

\section{Methods}
\label{bkm:Ref144506550}

Section \ref{bkm:Ref144496811} explains the corrosion dataset along with its seven variables.
Section \ref{bkm:Ref144507928} illustrates the arrangements and the necessary fine-tuning made to the dataset to use as input to the causal discovery algorithms. Section \ref{bkm:Ref144508337} explains the
significance levels and the assumptions made for analysing the causal graphs during the discussion. Sections \ref{bkm:Ref144508337} to
\ref{bkm:Ref144507965} explain the causal discovery algorithm and everything required for its implementation.

\subsection{Corrosion Data}
\label{bkm:Ref144496811}The corrosion dataset used in this work was formed by Dr Adriana Encinas-Oropesa in 2005,
during her Ph.D. thesis in collaboration with the Advanced Long Life Turbine Coating Systems project (ALLBATROS) \citep{encinas2005study}. To
produce the corrosion dataset, experiments were conducted on a single crystal CMSX-4. Three different metallic protective coatings, RT22, CN91 and LCO22 were applied to the CMSX-4
base alloy. The uncoated CMSX-4 material and the three coated versions of the CMSX-4 alloy were
treated as four different materials during the application of the causal discovery algorithm. The dataset consists of
two different operating temperatures, 700 and 900{\textdegree}C, with varying levels of gas compositions and deposit chemistries and fluxes. 

After an exposure time of 1000 hours, the material loss data were collected. This
data corresponded to three different flux deposition rates of 0.5, 1.5 and 5 {\textmu}g/cm$^2$/h. The gas composition consisted of a constant 300vppm (volumetric parts per million) of SO$_2$ along with
varying amounts of HCl (0 or 100vppm). Material loss values (due to hot corrosion) corresponding to each deposition
rate, temperature, material, and amount of gas composition were tabulated.

\subsubsection{Data Pre-Processing}
\label{bkm:Ref144507928}\label{bkm:Ref144508151} To asses the extent of hot corrosion, pre- and post-exposure sample dimensions were compared resulting in 30 values. The dataset formed gives material loss as a function of the cumulative probability of each amount of metal loss. The probability indicates the likelihood of having a certain amount of metal loss or more. Thus, the most extensive damage occurs with low probability. For simplicity, the cumulatively distributed dataset
was truncated to three values: highest, median, and lowest material loss. The highest and the lowest material loss (HML and LML) are used to
understand the factors that are dominantly influencing the hot corrosion at the opposite ends of the material loss
spectrum, while the median material loss (MML) helps in identifying the typical corrosion factors leading to hot
corrosion from an overall perspective.

`Amount of salt', `Temperature', `SO$_2$', `HCl', `Time
of exposure', `Material', and `Material Loss' were the seven variables tabulated. Although the variables could take continuous values, in the experiments they were fixed to specific values. Thus all are considered categorial variables for the causal discovery algorithm, except for the case of the Material Loss which took different, continuous values.

\subsection{Statistical Independence Tests}
\label{bkm:Ref144508337}

The cause-effect relationship between the
corrosion variables present in the dataset are analysed from a statistical point of view. The null hypothesis considered is a lack of any causal relationship. As such, results are presented with their significance level $\alpha $, which represents the probability of incorrectly rejecting the null hypothesis when it is true.  The confidence level (CL) follows the relation CL = 1-$\alpha $ \citep{LOFTUS2022137, banerjee2009hypothesis}.

Causal discovery methods use conditional independence tests (CIT) to identify causal links
between the variables of the dataset and attempt to eliminate the spurious correlations within those variables \citep{peters2017elements}. 
The independence between the set of nodes X and Y, conditional to a set of nodes Z, is written as X $\perp$ Y {\textbar} Z \citep{pearl2009causality}. The independence between variables can be inferred locally from the CIT, but also globally from the causal structures present in the graph and how they are connected. Analysing all the causal paths connecting any variables it can be inferred what is their causal relationship. A criterion commonly used in this regard is the d-separation of variables \citep{pearl2009causality}.

Causal discovery
algorithms use various statistical tests to assess the independence between variables, such as the Chi-squared $\chi^2$ test (non-parametric test measuring the goodness-of-fit between expected and observed frequencies which works well with large discrete categorical samples) \citep{cressie1989pearson,agresti2012categorical}, Fisher-Z test (used for partial and zero correlation, this parametric test assumes that
the variables are normally distributed and works mainly on large sample sizes of continuous variables) \citep{ramsey2014scalable,agresti2012categorical}, G-squared test (non-parametric test that highlights relationships between categorical variables with more than two levels) \citep{cressie1989pearson}, or the Kernel-based conditional independence test (KCI). In this work, it was used the latter.

\begin{figure}[!ht]
\noindent
\centerline{ \includegraphics[width=0.5\textwidth]{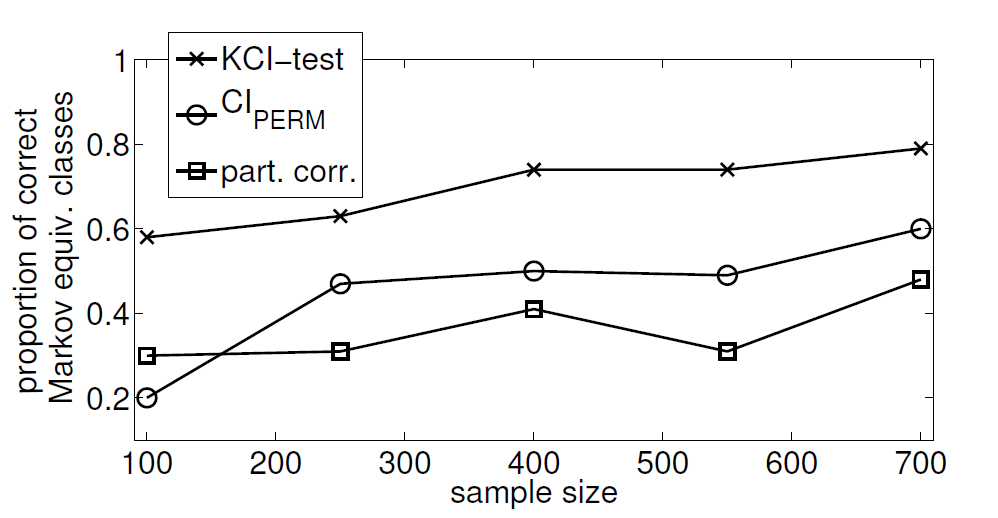} }
\caption{\label{bkm:Ref144507387}Accuracy of different CITs to infer the correct Markov Equivalence Class as a function of the sample size \citep{zhang2012kernel}.}
\end{figure}

The KCI test is a non-parametric test that can be derived from the kernel matrices of the variables under consideration, which characterize the similarity of the samples of those variables \citep{zhang2012kernel,yang2021model}. These kernel functions recognise non-linear
relationships between data points. This test can be applied to discrete or continuous
variables. Figure \ref{bkm:Ref144507387} shows the evolution of the accuracy of the test as a function of the number of samples for the dataset analysed in the original publication of the method.

\subsection{Causal Links, Structures and Graphs}
\label{bkm:Ref144508580}The different causal nature of the relationships between the variables can be represented with different types of causal links \citep{hasan2023survey}, as illustrated in Table \ref{bkm:Ref144507456}. The relationship depends on the variables measured and on possible confounding variables (those unmeasured
variables which influence the underlying causal mechanism)
\citep{nogueira2022methods}.

\begin{table*}[!ht]
    \caption{\label{bkm:Ref144507456}Different types of causal links and their respective denotations ignoring any selection bias \citep{nogueira2022methods,zhang2008completeness,zhang2008causal} }
    \centering
    \begin{tabular}{ccc}
    \hline
        \textbf{Causal Link} & &  \textbf{Description}  \\ \hline
        A $\rightarrow $ B (directed) & & A is the cause of B  \\ 
        A ― B (undirected) & &  Undetermined. A can cause B and B can cause A   \\ 
        A $\leftrightarrow$ B       (bidirected) & & \makecell{A and B do not cause each other but have a latent common cause.\\Confounding variables between A and B} \\ 
         A o$\rightarrow $ B (partially directed)  & A $\rightarrow $ B & A causes B (“o” turns into tail end)\\ 
         & A $\leftrightarrow$ B & There exists a confounder between A and B (“o” turns into arrow end)  \\ 
        A o―o B (undirected with “o” ends)  & A $\rightarrow $ B & A causes B  \\
         & A $\leftarrow $ B & B causes A  \\
         & A $\leftrightarrow$ B & A and B do not cause each other. Confounder between A and B  \\
         &  & This last option can be also combined with the previous two  \\
    \end{tabular}
\end{table*}

After applying the CIT to the variables from the dataset, the result of the causal discovery algorithm is a graphical
representation of the causal links between the variables called a causal graph \citep{hasan2023survey}. Chains, forks, and colliders (also called v-structures) are the three building blocks used in these causal graphical models to illustrate the cause-effect relationship
between the variables.

\begin{figure}[!ht]
\noindent
 \centerline{\includegraphics[width=0.5\textwidth]{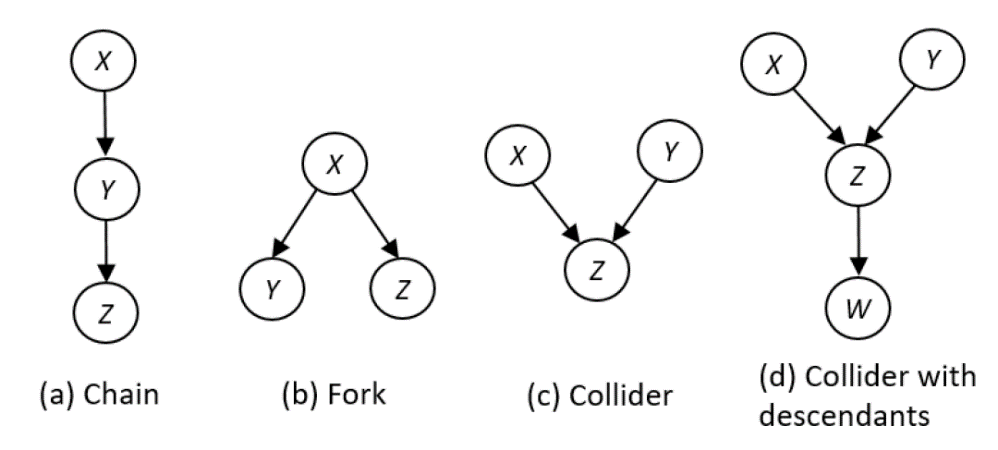} }
\caption{\label{bkm:Ref144507516}Building blocks of a Causal Graph \citep{hasan2023survey}}
\end{figure}

Figure \ref{bkm:Ref144507516}(a) shows the chain structure wherein X $\rightarrow $ Y $\rightarrow $ Z forms a chain where X causes Y and Y causes Z, therefore the conditional independence can be written as X $\perp$ Z
{\textbar} Y \citep{pearl2016causal}. Figure \ref{bkm:Ref144507516}(b) shows the fork structure Y $\leftarrow $ X $\rightarrow $ Z wherein the
node X forms a directed edge towards Y and Z making it the common ancestor for Y and Z \citep{pearl2016causal}. Since
there is only one path between Y and Z through X based on the d-separation criterion, conditional independence can be
written as Y $\perp$ Z {\textbar} X. Figure \ref{bkm:Ref144507516}(c) shows a collider X $\rightarrow $ Z $\leftarrow $ Y, wherein the descendant Z has two common ancestors X and Y. Even though there is one path between X and Y, the presence of a collision node Z makes X and Y
conditionally dependent given the collision node Z, i.e. X $\not\perp$ Y {\textbar} Z. Figure \ref{bkm:Ref144507516}(d) shows a
collider structure with extra descendant W where X $\not\perp$ Y {\textbar} Z and X $\not\perp$ Y {\textbar} W.

A causal directed acyclic graph (DAG) consists of a set of random variables with edges between them which never form a directed cycle within the graph \citep{pearl2009causality}. Generally, causal discovery algorithms do not allow identifying the causal graph, but the Markov Equivalence Class (MEC) of the graph. If two DAGs are Markov equivalent, they have the same skeleton and set of colliders as well as the same (conditional) independencies \citep{peters2017elements}, which is what the algorithms can usually identify. Once the class is identified, using interventions in the variables it is then possible to identify the actual causal DAG \citep{he2015counting}.
 
\begin{figure}[!ht]
\noindent
 \centerline{\includegraphics[width=0.5\textwidth]{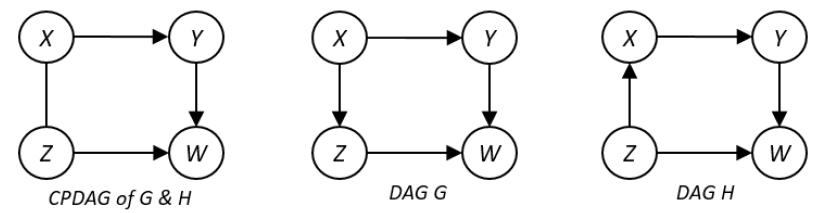} }
\caption{\label{bkm:Ref144507608}Example of Markov Equivalence Class \citep{hasan2023survey}
}
\end{figure}

The example taken from \citep{hasan2023survey} in Figure \ref{bkm:Ref144507608} shows a Completed Partially DAG (CPDAG) of G and H which represents the union of the Markov
equivalent DAGs G and H. The undirected edge between X and Z in the CPDAG suggests that it might contain X $\rightarrow $ Z
(shown in the DAG G) or Z $\rightarrow $ X (shown in the DAG H).

Table \ref{bkm:Ref144507624}, shows the different edges that can be observed in each type of causal graph. Each causal discovery
algorithm produces as an output a different type of causal graph.

\begin{table*}[!ht]
    \caption{\label{bkm:Ref144507624}Causal Graphs and their types of edges \citep{richardson2002ancestral, perkovic2020identifying, tian2012generating, hu2020faster,hasan2023survey}}
    \centering
    \begin{tabular}{ccccc}
    \hline
        ~ & \textbf{Directed} ($\rightarrow$) & \textbf{Undirected }(—) & \textbf{Bi-directed} ($\leftrightarrow$) & \textbf{Partially Directed} (o$\rightarrow$) \\ \hline
        DAG & X & ~ & ~ & ~ \\ 
        Partially DAG (PDAG) & X & X & ~ & ~ \\ 
        Completed PDAG (CPDAG) & X & X & ~ & ~ \\ 
        Maximal Ancestral Graph (MAG) & X & X & X & ~ \\ 
        Partial Ancestral Graph (PAG) & X & X & X & X \\
    \end{tabular}
\end{table*}

\subsection{Causal Discovery Algorithms}
\label{algos}The two main categories of causal discovery algorithms are constraint-based and score-based methods. The first checks the graph structure against the independence constraints imposed by the data. In the second method, possible graphs are scored for their ability to fit the data. In the latter, the space of DAGs is searched to find the graph that maximises the score. This last method is especially useful when dealing with a large number of variables since the combinatorial space of possible graphs grows exponentially. In this work, since a small number of variables are studied it is used a constraint-based method. 

The constraint-based algorithms use CITs to investigate the type of edges between
the variables or their absence \citep{glymour2019review}. One of the earliest and most common of these algorithms is the PC (Peter \& Clark) Algorithm \citep{spirtes2000causation}. It uses CITs to understand the underlying causal
mechanism of the causal structures. It assumes independent and identically distributed (i.i.d)
samples and absence of confounding variables.

Here it is used the Fast Causal Inference (FCI) algorithm \citep{spirtes2000causation}. It is a variant of the PC algorithm that provides asymptomatically
correct results while considering the presence of confounding variables in a dataset with i.i.d samples \citep{glymour2019review}. The output causal graph of the FCI algorithm is a Partial Ancestral Graph (PAG), including the
presence of directed, undirected, partially directed and bi-directed edges \citep{hasan2023survey}.

\subsection{FCI Algorithm Implementation}
\label{bkm:Ref144507965}The causal-learn\footnote{~\url{https://github.com/py-why/causal-learn}} package \citep{zheng2023causal} was used for
implementing the FCI algorithm. The dataset used as the input of the FCI
algorithm included the variables shown in Table \ref{bkm:Ref144508200}.

\begin{table*}[!ht]
    \caption{\label{bkm:Ref144508200}Categories of variables present in the datasets for highest (HML), median (MML) and lowest (LML) material loss}
    \centering
    \begin{tabular}{cccccccccccc}
    \hline
        \textbf{Mat. Loss / Dataset} & \multicolumn{4}{c}{\textbf{Materials}}  & \multicolumn{2}{c}{\textbf{Temperature (\textdegree C})}  & \multicolumn{3}{c}{\textbf{Amount of Salt (µg/cm$^2$/h})}  &  \multicolumn{2}{c}{\textbf{HCl (ppm)}}  \\ \hline
        HML & 1 & 2 & 3 & 4 & 700 & 900 & 0.5 & 1.5 & 5.0 & 0 & 100 \\ 
        MML & 1 & 2 & 3 & 4 & 700 & 900 & 0.5 & 1.5 & 5.0 & 0 & 100 \\ 
        LML & 1 & 2 & 3 & 4 & 700 & 900 & 0.5 & 1.5 & 5.0 & 0 & 100 \\ 
    \end{tabular} 
\end{table*}

From the initial seven variables available, mentioned in section \ref{bkm:Ref144508151}, SO$_2$ and Time of Exposure were removed from the dataset because of their constant values. The rest of the variables were used as input to the
algorithm. The four materials used in the experiments were assigned numbers from 1 to 4, respectively. The algorithm was applied with the default parameter settings, CIT = KCI and no background
knowledge. Since the dataset consisted of only five variables, no background knowledge was
introduced into the causal algorithm to avoid selection- or expert-bias. Therefore, the causal graphs formed were
purely based on the observational data. A range of 1\% to 99\% significance level, with a 1\% step increment, was implemented to observe how the causal links would form
with high and low confidence levels.

\section{Results}
\label{bkm:Ref144506558} This section presents the causal graphs obtained by implementing the FCI
algorithm on the material loss datasets. The descriptions of the algorithms, graphs and types of links can be found in Sections \ref{bkm:Ref144508580} and \ref{algos}.

Based on the degree of material loss, the dataset was divided into
three parts: Highest Material Loss (HML), Median Material Loss (MML) and Lowest Material Loss (LML) (see Table \ref{bkm:Ref144508200}). Varying significance levels from 1\% to 99\% were considered in the FCI algorithm.
The results are differentiated according to HML, MML and MML and the significance level of each graph. 

Table \ref{kci_fci_hml} shows the causal graphs obtained by applying KCI in the FCI algorithm on the HML dataset. The ``o'' termination in the
causal link formed between Amount of Salt and Material Loss at  $\alpha=1\% $, illustrates that it can be either an
arrowhead ({\textgreater}) or a tail end of a directed edge (see Table \ref{bkm:Ref144507456}). Hence, the direction of the causal relationship is not
clearly depicted at $\alpha=1\% $ between the two variables. 

At  $\alpha=7\% $, partially directed edges ``o\-\-$\rightarrow $'' from Amount of Salt and HCl to Material Loss were
formed. This means that if ``o'' becomes an arrowhead, it forms a bidirected edge. Table \ref{bkm:Ref144507456} shows that the bidirected
edge indicates that there is an unmeasured confounder present between the two variables. On the other hand, if ``o''
becomes a tail end, it confirms that Amount of Salt and HCl causes Material Loss. Similarly, Temperature and Material
formed partially directed edges with Material Loss at $\alpha=8\% $ and $\alpha=60\% $  for HML conditions, respectively.

\begin{table*}[!ht]
    \caption{\label{kci_fci_hml}Causal graphs using KCI in FCI algorithm with increasing significance levels for HML dataset}
    \centering
    \begin{tabular}{cc}
    \hline
        $\mathbf{\alpha}$ & \textbf{Causal Graphs} \\ \hline
        0.01 & ~  \includegraphics[width=0.8\textwidth]{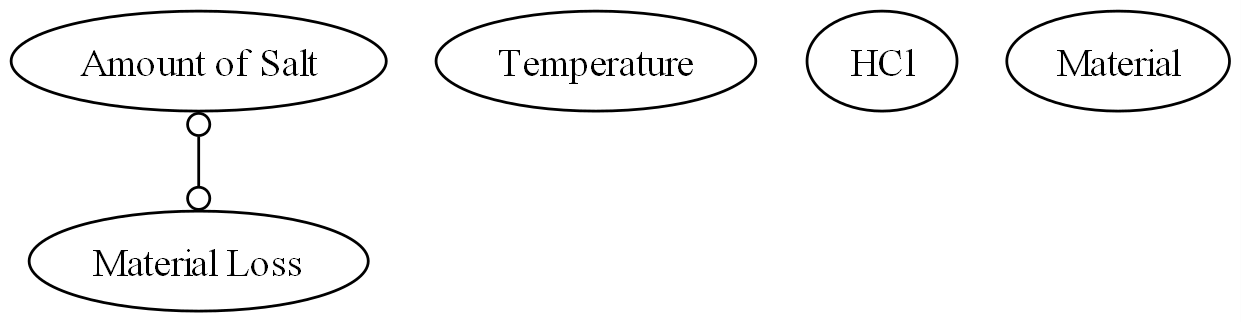} \\ 
        0.07 & ~  \includegraphics[width=0.8\textwidth]{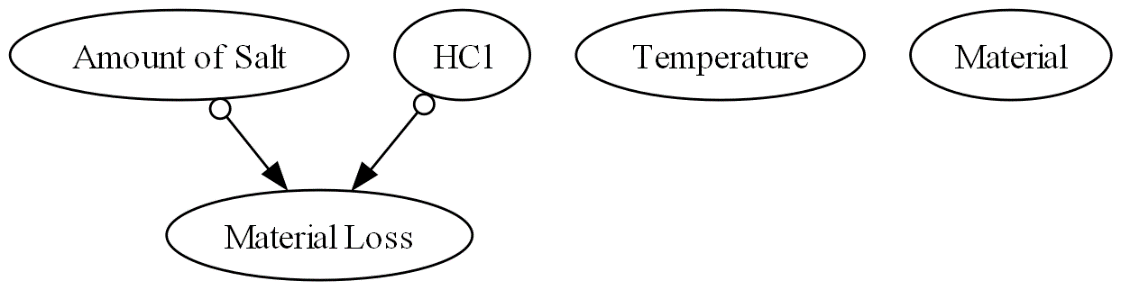} \\ 
        0.08 & ~  \includegraphics[width=0.8\textwidth]{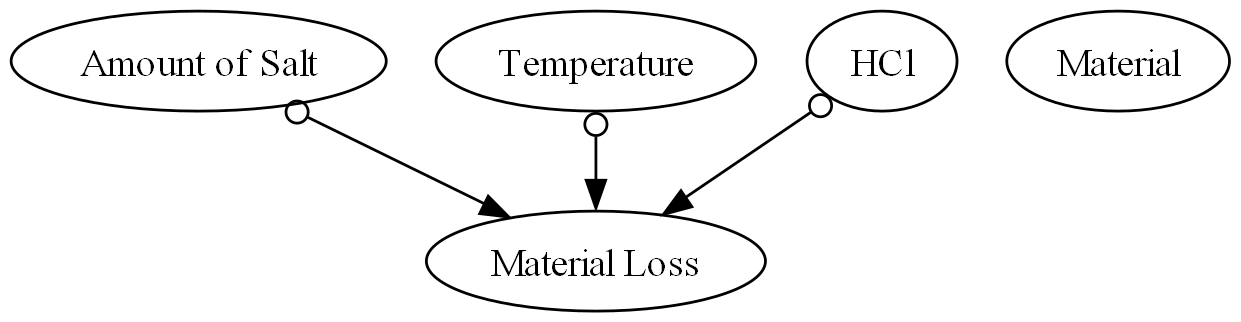} \\ 
        0.6 & ~  \includegraphics[width=0.8\textwidth]{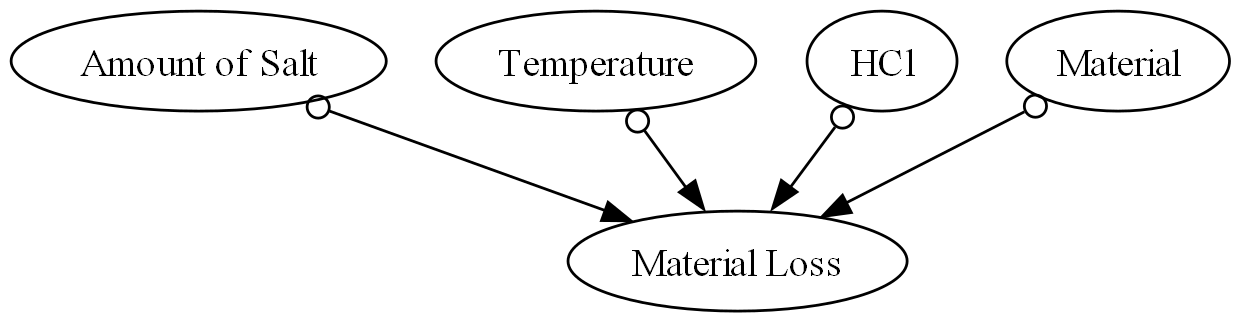} \\ 
    \end{tabular}
\end{table*}

Table \ref{kci_fci_mml} shows the results for the MML dataset. 
The undirected edges with ``o'' ends which formed between Amount of Salt and Material Loss at  $\alpha=1\% $, get converted to a partially directed edge from Amount of Salt to Material Loss, along with a
partially directed edge from Temperature to Material Loss at  $\alpha=9\% $. Gradually, partially directed edges were
formed from HCl and Material to Material Loss at $\alpha=20\%  $ and $\alpha=60\%  $, respectively.

\begin{table*}[!ht]
    \caption{\label{kci_fci_mml}Causal graphs using KCI in FCI algorithm with increasing significance levels for MML dataset}
    \centering
    \begin{tabular}{cc}
    \hline
        $\mathbf{\alpha}$ & \textbf{Causal Graphs} \\ \hline
        0.01 &
 \includegraphics[width=0.8\textwidth]{Figure_7.png} \\
0.09 &
 \includegraphics[width=0.8\textwidth]{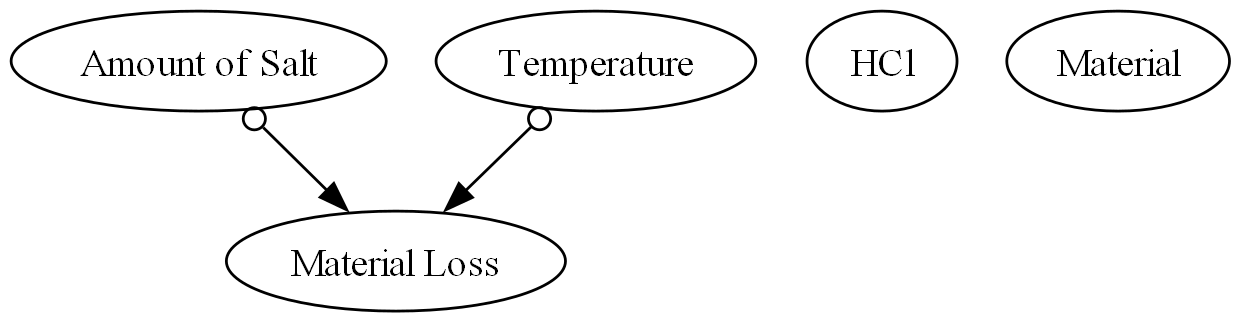} \\
0.2

 &
 \includegraphics[width=0.8\textwidth]{Figure_9.png} \\
0.6 &
 \includegraphics[width=0.8\textwidth]{Figure_10.png} \\
     \end{tabular}
\end{table*}

The causal graphs for the LML dataset are presented in Table \ref{kci_fci_lml}. 
At $\alpha=1\%$ for the LML dataset, an undirected edge with ``o'' ends was formed between Amount of Salt and Material
Loss. With the increase in $\alpha $ to 20\%, 50\% and 70\%, partially directed edges were formed
from Amount of Salt and Temperature, HCl and Material to Material Loss, respectively.

\begin{table*}[!ht]
    \caption{\label{kci_fci_lml}Causal graphs using KCI in FCI algorithm with increasing significance levels for LML dataset}
    \centering
    \begin{tabular}{cc}
    \hline
        $\mathbf{\alpha}$ & \textbf{Causal Graphs} \\ \hline
0.01 &
 \includegraphics[width=0.8\textwidth]{Figure_7.png} \\
0.2 &
 \includegraphics[width=0.8\textwidth]{Figure_11.png} \\
0.5 &
 \includegraphics[width=0.8\textwidth]{Figure_9.png} \\
0.7 &
 \includegraphics[width=0.8\textwidth]{Figure_10.png} \\
     \end{tabular}
\end{table*}

\section{Discussion}
\label{bkm:Ref144506571} In this section it is discussed the nature of the current corrosion dataset and its drawbacks, as well as the causal graphs
formed by the FCI algorithm and their key findings. 

The type of dataset that is used as an input to the algorithm is one of the major factors
determining the result. The current corrosion dataset was designed to maintain control over the corrosion variables to
understand the effects on the material loss. The causal graphs in Tables \ref{kci_fci_hml}to \ref{kci_fci_lml},  show that all the corrosion variables are
directed towards Material Loss with increasing significance values. This means that the variables were controlled to
observe the amount of material loss in the base material. However, for a pure causal inference study, randomisation in the
variables would have offered a greater benefit \citep{spirtes2000causation}. For instance, if all the variables were uncontrolled and
randomised, observing causal links within the variables could have been a possibility. This would have helped in
understanding the cause-effect relationship between the variables and not just through the Material Loss.  The degree of influence that each variable has on Material Loss with varying significance levels can be observed in Tables \ref{kci_fci_hml} to \ref{kci_fci_lml}. However, no claims can be made about the
presence of causal relationships between `Amount of Salt', `Temperature', `HCl', and `Material' using the current
dataset.

As discussed in sections \ref{bkm:Ref144509428} and \ref{bkm:Ref144509436}, the salt deposition on the material surface
is a significant factor in causing hot corrosion and, indeed, hot corrosion is also defined as a deposit-induced accelerated form
of corrosion \citep{pettit2011hot}. The salt deposits initiate the breakdown of the protective oxide layer of the substrate
which is the starting point of hot corrosion \citep{eliaz2002hot}. Therefore, even at smaller significance values,
Amount of Salt makes a directed or an undirected causal link with Material Loss for all the material loss levels (HML, MML, and LML). 

Operating temperature facilitates the corrosive environment that enables hot
corrosion \citep{pettit2011hot}. Moreover, the lower melting point of the eutectic mixtures formed by different salts
accelerates the corrosion process \citep{eliaz2002hot}, hence proving that temperature is another factor that has a
significant influence on hot corrosion. Indeed, for MML and LML temperature becomes the second
significant variable that influences hot corrosion in the given dataset, at smaller significance values. For MML and LML HCl in the gas only
appears to form a causal link with Material Loss at higher significance levels. Chloride-contaminated oxide layers are known to accelerate the corrosion rate \citep{viklund2011hcl}. HCl is a crucial factor that participates in increasing hot
corrosion rates. More experimental
research is required to clarify the influence of HCl. However, this analysis shows its significance for driving extreme metal loss such as pitting.

Table \ref{kci_fci_hml}, formed using the HML dataset, shows that at 1\%
significance level, the outcome from FCI shows an undirected causal link between Amount of Salt and Material Loss with
``o'' at either end. This means that at such a conservative significance value, FCI is only able to infer the existence
of a link between the two variables without pointing out the cause-effect relationship. The undirected causal link can
also point towards the probability of incorrectly accepting the null hypothesis i.e., Type-II error. The ``o'' ends can
be perceived as an arrowhead or a tail end of a directed edge because the algorithm is too conservative in assigning
them to form a meaningful causal link. In case both ``o'' ends become arrowheads, then it means that there is a
confounding variable that has not been considered in the dataset. 

In Table \ref{kci_fci_hml}, with a significance value of 7\%, the three partially directed (o$\rightarrow $) causal
links confirm that there is a possibility of Amount of Salt, HCl, and Temperature being factors causing
Material Loss, or there exist unmeasured confounders between them that are influencing the behaviour of those variables. The FCI algorithm is not able to clarify the causal relationship between these variables and shows the
possibility that confounder variables are present. HCl forms the same partially directed edge with Material Loss and a similar
argument can be made for this causal link as well. Under the specific conditions of the test, it seems that HCl is linked to HML, so it seems as HCl is one of the driving forces involved in the formation of pitting. This provides a very useful insight into the mechanism of hot corrosion, as the cause of pit initiation is still under debate. From  $\alpha=7\%$,  Temperature also forms a partially
directed causal link with Material Loss. This suggests that for the HML dataset, HCl and Temperature are the most
influential factors advancing material loss, after Amount of Salt. There is also a possibility that unmeasured
confounders exist which influence all the said variables. Therefore, datasets with a larger sample size and greater
number of variables can further help in understanding the presence of these confounders. 

Table \ref{kci_fci_mml} shows that the FCI algorithm is not confident enough in forming causal links between the variables up to 9\%
significance value, for the MML dataset. At this value, both Amount of Salt and Temperature form partially directed causal
links pointing towards Material Loss. Amount of Salt and Temperature are observed to be the
dominant factors accelerating the Material Loss because the salt deposits initiate the breaking of the protective layer
of the base material and high operating temperature facilitates such corrosion mechanisms. There is a possibility that
the inclusion of variables like partial pressure would have addressed the presence of the unmeasured confounder at LTHC
(700{\textdegree}C). This is because LTHC is a function of temperature and partial pressure, as discussed in section
\ref{bkm:Ref144509601}. 

For the LML condition in Table \ref{kci_fci_lml}, the undirected causal link with ``o'' ends formed between Amount
of Salt and Material Loss at 1\% significance value suggests that at such a low level of material loss, Amount of Salt
might have just started to break down the protective layer to cause the Material Loss. Or there is an unmeasured
variable that exists which is influencing both variables.

Results beyond the 10\% significance level are not discussed since they are not statistically significant.

The causal links shown in the previous tables suggest the possible presence of confounding variables. This opens up the avenue
to design experiments that also cater to other variables that were not included in this dataset. The FCI algorithm
 indicates that for the experimental conditions represented in this study's limited dataset, Amount of Salt has the maximum influence on Material Loss.

\section{Conclusions}
\label{bkm:Ref144506584}
Understanding the underlying causal mechanisms between the factors leading to hot corrosion in gas turbines can
highlight the inner workings of this process. Implementing causal discovery methods can help in recognising the causal relationships between these
corrosion factors. These methods were applied to three different datasets that were divided based on the degree of
material loss observed on the materials tested: highest, median, and lowest material loss. The
causal discovery algorithm FCI was applied to produce the causal graphs that illustrate the causal relationships between
the corrosion variables present in these three datasets. A wide range of significance levels was analysed, to showcase
the confidence level with which the causal relationships were formed between the corrosion variables.

After analysing the causal graphs for the given range of significance levels, it was observed that the number of causal
links decreased in the order of highest to lowest material loss. As the degree of material loss decreased to the median range, only two causal
links i.e., Amount of Salt and Temperature to Material Loss were formed within the set range of
significance levels. Eventually, only a single undirected causal link was formed between Amount of Salt and Material
Loss for LML conditions. This showcased that there is uncertainty in claiming the real causal relationship between the
two variables at low degrees of material loss. Causal graphs produced using the FCI algorithm suggested the possible
existence of unmeasured confounding variables.

From this nascent-stage study of causality in hot corrosion, it can be concluded that Amount of Salt and Temperature
were the typical factors causing Material Loss from an overall perspective. However, HCl also proved to be a dominant factor for the
HML dataset. As this dataset can include pitting regions this can give insight into factors driving pit formation rather than more average metal loss.

The FCI algorithm proved beneficial in understanding the causal relationships, but a randomised and uncontrolled type
of corrosion dataset can further help in future causal research.

Since this study is at a nascent phase of its research timeline, several avenues for future research can be delved into
to improve the understanding and application of causal discovery techniques. Following are the recommendations that can
potentially improve the quality of further studies:
\begin{enumerate}
\item A randomised dataset which can help in performing a more comprehensive causal discovery study should be produced. 
\item Including variables such as partial pressures of the gaseous contaminants, crystalline structures, and varying gas
compositions can improve the depth of the dataset.
\item The sample sizes must be increased to
generate stronger causal links with higher confidence levels.
\item If a time-dependent dataset is produced, causal discovery algorithms for
time series data can be implemented such as Granger Causality-based algorithms \citep{hasan2023survey}. 
\end{enumerate}
These recommendations can help in designing the next phase of the causal discovery study of hot corrosion. Expanding
upon this study can further help in developing predictive maintenance and material degradation detection models.

\subsection*{Data Availability Statement}
The data that support the findings of this study are available from the corresponding author upon reasonable request.

\bibliography{causal-corrosion}

\end{document}